\documentclass[aps,prb,amsmath,amssymb,superscriptaddress,twocolumn,longbibliography]{revtex4-2}
\usepackage{bbm}
\usepackage{graphicx}% Include figure files
\usepackage{dcolumn}% Align table columns on decimal point
\usepackage{bm}% bold math
\usepackage{subfigure}
\usepackage{amsmath}
\usepackage{stmaryrd}
\usepackage{times}
\usepackage{graphicx}
\usepackage{dcolumn}
\usepackage{color}
\usepackage{makecell}
\usepackage{tabularx,multirow}
\usepackage{amssymb}% http://ctan.org/pkg/amssymb
\usepackage{pifont}% http://ctan.org/pkg/pifont
\usepackage[colorlinks,linkcolor=blue,citecolor=blue,urlcolor=black]{hyperref}
\begin{document}
\title{Anisotropic nonlinear transport in two-dimensional ferroelectrics}
\author{Qin Zhang}
\affiliation{School of Physics, Harbin Institute of Technology, Harbin 150001, China}
\author{Xu Chen}
\affiliation{School of Physics, Harbin Institute of Technology, Harbin 150001, China}
\author{Mingbo Dou}
\affiliation{School of Physics, Harbin Institute of Technology, Harbin 150001, China}
\author{M. Ye. Zhuravlev}
\affiliation{St. Petersburg State University, St. Petersburg 190000, Russia}
\author{A. V. Nikolaev}
\affiliation{Skobeltsyn Institute of Nuclear Physics, Moscow State University, Moscow 101000, Russia}
\author{Xianjie Wang}
\email{Contact author: wangxianjie@hit.edu.cn}
\affiliation{School of Physics, Harbin Institute of Technology, Harbin 150001, China}
\affiliation{Frontiers Science Center for Matter Behave in Space Environment, Harbin Institute of Technology, Harbin 150001, China}
\affiliation{Heilongjiang Provincial Key Laboratory of Advanced Quantum Functional Materials and Sensor Devices, Harbin 150001, China}
\author{L. L. Tao}
\email{Contact author: lltao@hit.edu.cn}
\affiliation{School of Physics, Harbin Institute of Technology, Harbin 150001, China}
\affiliation{Frontiers Science Center for Matter Behave in Space Environment, Harbin Institute of Technology, Harbin 150001, China}
\affiliation{Heilongjiang Provincial Key Laboratory of Advanced Quantum Functional Materials and Sensor Devices, Harbin 150001, China}
\date{\today}
\begin{abstract}
The longitudinal nonlinear response plays a crucial role in the nonreciprocal charge transport and may provide a simple electrical means to probe the spin-orbit coupling, magnetic order and polarization states, etc. Here, we report on a study on the polarization and magnetic field control of longitudinal nonlinear transport in two-dimensional (2D) ferroelectrics with in-plane polarization. Based on the Boltzmann transport theory, we first study that using a general Hamiltonian model and show that the nonlinear conductivity can be significantly tuned by the polarization and magnetic field. In addition, the nonlinear conductivity reveals a strong spatial anisotropy. We further derive the analytical formulas for the anisotropic nonlinear conductivity in exact accordance with numerical results. Then, we exemplify those phenomena in the 2D ferroelectric SnTe monolayer  in the presence of an external magnetic field based on the density functional theory calculations. It is also revealed that the polarity of nonlinear conductivity is locked to the direction of the polarization, thus pointing to the possibility of the nonlinear detection of polarization states. Our work uncovers intriguing features of the longitudinal nonlinear transport in 2D ferroelectrics and provides guidelines for designing the polarization control of rectifying devices.
\end{abstract}
\maketitle
\section{Introduction}
The nonlinear transport refers to the charge or Hall current contributed from the second (high) order in an electric field and can be classified into longitudinal (charge, dissipative) and transverse (Hall, non-dissipative) nonlinear responses\cite{Callawaybook,sae1602390,nc3740,nm2025}. For example, it was predicted that a nonlinear anomalous Hall effect can be induced by the Berry curvature dipole\cite{prl216806,nrp744} in $\mathcal{T}$-invariant but $\mathcal{P}$-broken ($\mathcal{T}$ for time-reversal and $\mathcal{P}$ for inversion) systems\cite{prb041101}. Later, this effect was experimentally confirmed in Weyl semimetals WTe$_2$ and MoTe$_2$\cite{nature337,nm324,nc2049}. In addition to the Berry curvature dipole, the nonlinear anomalous Hall effect can also be caused by the quantum metric\cite{prl166601}, which characterizes the quantum geometry of wave functions\cite{rmd1959,Wilczek} and has been demonstrated in compensated antiferromagnets such as CuMnAs\cite{prl277201} and Mn$_2$Au\cite{prl277202}.

On the other hand, the longitudinal nonlinear response is responsible for the nonreciprocal charge transport (NCT), which is characterized by unequal resistances $R$ for a material with opposite currents $I$\cite{sae1602390,nc3740,nrp558}, that is, $R(+I)\neq R(-I)$. It is commonly held that the NCT effect arises from the nonlinear Drude conductivity caused by a band asymmetry\cite{sae1602390,nc3740}. Indeed, the NCT effect has been observed in some noncentrosymmetric materials by applying magnetic fields\cite{np578,nc540,prr033253,prb115202,prl176602,prl046303,prr013041,prb155411}, ferromagnetic\cite{prb054429} and antiferromagnetic\cite{prl276601,na487,prl096802,prl046801} materials due to the combined spin-orbit coupling (SOC) and magnetic order. In addition to the Drude mechanism, the longitudinal nonlinear conductivity can also be contributed by the quantum metric\cite{prl026301,prbL201405}. As distinct from the nonlinear Drude conductivity proportional to $\tau^2$ ($\tau$ for relaxation time), the quantum metric induced nonlinear conductivity is $\tau$ independent and thus represents an intrinsic contribution\cite{na487,prl026301}.

As distinct from the linear conductivity, the nonlinear conductivity represents a $\mathcal{P}$-odd quantity, whose polarity is reversed under $\mathcal{P}$ operation\cite{Callawaybook}. In ferroelctrics, switching polarization is equivalent to $\mathcal{P}$ operation\cite{jpd113001}, which in turn reverses the polarity of the nonlinear conductivity. Thus, it is instructive to explore the polarization control of the nonlinear transport in ferroelctrics. Very recently, Kondo \emph{et al}\cite{prr013041} investigated the NCT in the polar Dirac metals BaMnX$_2$ (X=Sb, Bi) in the presence of magnetic fields. It was found that the nonreciprocal resistivity manifests (vanishes) when the current is perpendicular (parallel) to the polarization consistent the selection rule for the nonreciprocal transport\cite{nc3740}. However, the systematic study on the interplay between the nonlinear conductivity, polarization and magnetic field have not been well explored. In particular, the spatial anisotropy and analytical formulas for the nonlinear conductivity in 2D ferroelectrics remain to be explored. 

Here, we report on the theoretical study on the polarization and magnetic field tunable nonlinear transport in 2D ferroelectrics based on the Boltzmann transport theory. We discuss its spatial anisotropy and derive the analytical formulas for the nonlinear conductivity to clarify the interplay between the nonlinear conductivity, polarization and magnetic field. 

The rest of the paper is organized as follows. In Sec. \ref{sec2}, we present the theoretical formalism and computational details for the nonlinear conductivity calculations. In Sec. \ref{sec3}, we discuss the anisotropic nonlinear transport in 2D ferroelectrics based on the general Hamiltonian model results in Sec. \ref{subA} and DFT results for the 2D ferroelectric SnTe monolayer in Sec. \ref{subB}. Finally, Sec. \ref{sec4} is reserved for further discussion and conclusion.

\section{Theoretical formalism and computational details\label{sec2}}

\begin{figure}
\includegraphics[width=0.4\textwidth]{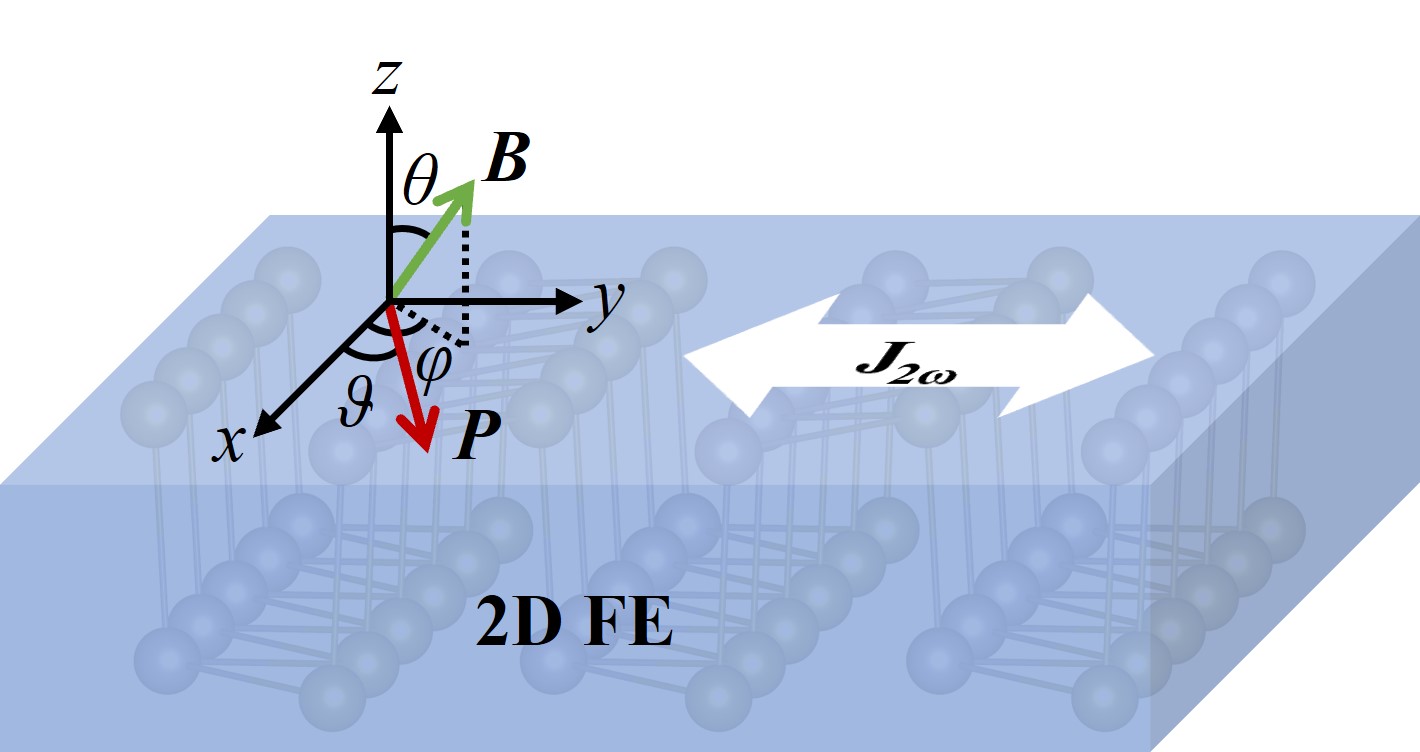}%
\caption{\label{f-1} Schematic illustration of 2D ferroelectrics with in-plane polarization $\mathbf{P}$ and an external magnetic filed $\mathbf{B}$. The directions of $\mathbf{B}$ and $\mathbf{P}$ are denoted by the unit vectors $\hat{\mathbf{B}}=(\sin\theta\cos\varphi, \sin\theta\sin\varphi, \cos\theta)$ and $\hat{\mathbf{P}}=(\cos\vartheta, \sin\vartheta, 0)$, respectively. $\theta$ denotes the polar angle while $\varphi$ and $\vartheta$ represent the azimuthal angles. $\mathbf{J}_{2\omega}$ represents the second-harmonic current  at frequency $\omega$ due to the nonlinear response.}
\end{figure}

To second order in an applied electric field $\mathbf{\mathcal{E}}$, the produced current density $\mathbf{J}$ in a solid is given by\cite{Callawaybook}
\begin{equation}\label{eq-1}
  J_a=\sigma^{(1)}_{ab}\mathcal{E}_b+\sigma^{(2)}_{abc}\mathcal{E}_{b}\mathcal{E}_{c},
\end{equation}
where $\sigma^{(1)}$ and $\sigma^{(2)}$ are the first-order linear and second-order nonlinear conductivities, respectively. For 2D systems, the indices $a, b, c=x, y$ denote Cartesian components and a summation over repeated indices is implied. Here we consider the Drude conductivity and, under the relaxation time $\tau$ approximation, $\sigma^{(1)}_{ab}$ and $\sigma^{(2)}_{abc}$ for two dimensions take the forms\cite{Callawaybook,prb054429,prl096802,prb224423}
\begin{equation}\label{eq-2}
    \sigma^{(1)}_{ab} = \frac{e^2\tau}{4\pi^2\hbar^2}\sum_n\int f_n\frac{\partial^2\epsilon_{n\mathbf{k}}}{\partial k_a\partial k_b}d^2k,
\end{equation}
and
\begin{equation}\label{eq-3}
    \sigma^{(2)}_{abc} = -\frac{e^3\tau^2}{4\pi^2\hbar^3}\sum_n\int f_n\frac{\partial^3\epsilon_{n\mathbf{k}}}{\partial k_a \partial k_b \partial k_c}d^2k,
\end{equation}
where $f_n(\epsilon_{n\mathbf{k}}, \epsilon_F)$ is the Fermi distribution function given in terms of the eigenvalue of the $n$th band $\epsilon_{n\mathbf{k}}$ and the Fermi energy $\epsilon_F$. 

On the other hand, it is instructive to examine the spatial anisotropy of the longitudinal nonlinear conductivity, that is, $a=b=c$ in Eq. \ref{eq-3}. For $\mathbf{\mathcal{E}}$ along the $\phi$ direction ($\phi$ for azimuthal angle), with the aid of directional derivatives and the symmetry of second derivatives, one finds
\begin{widetext}
\begin{equation}\label{eq-4}
\sigma^{(2)}_{\phi\phi\phi}=\sigma^{(2)}_{xxx}\cos^3\phi+\sigma^{(2)}_{yyy}\sin^3\phi+(\sigma^{(2)}_{xyy}+2\sigma^{(2)}_{yxy})\sin^2\phi\cos\phi+(\sigma^{(2)}_{yxx}+2\sigma^{(2)}_{xxy})\sin\phi\cos^2\phi.
\end{equation}
\end{widetext}

Our DFT calculations were performed using the plane-wave ultrasoft pseudopotential method\cite{prb7892} as implemented in the QUANTUM ESPRESSO\cite{jpcm395502,jpcm465901,jcp154105}. For the SnTe monolayer, an energy cutoff of $50$ Ry for the plane-wave expansion, a $k$-point mesh of $10\times10\times1$ for the Brillouin zone integration and generalized gradient approximation (GGA)\cite{prl3865} for the exchange and correlation functional were adopted. The atomic coordinators were fully relaxed with the force tolerance of $10^{-4}$ Ry/Bohr. A vacuum region of more than $20$ Å along the $z$ direction was imposed. The conductivities $\sigma^{(1)}_{ab}$ and $\sigma^{(2)}_{abc}$ were calculated by using the Boltzmann transport theory under relaxation time $\tau$ approximation and the Hamiltonian was constructed from the maximally-localized Wannier function basis set\cite{prb12847,rmp1419} as implemented in the Wannier90 code\cite{jpcm165902}. We used the temperature of $300$ K in the Fermi distribution function and the $k$-point mesh of $501\times501\times1$ for conductivity calculations.

\section{Anisotropic nonlinear transport\label{sec3}}

In this section, we investigate the anisotropic nonlinear transport by examining the magnetic field and polarization tunable $\sigma^{(2)}_{\phi\phi\phi}$. We present the general model results in Sec. \ref{subA} and DFT results for the 2D ferroelectric SnTe monolayer in Sec. \ref{subB}.

\subsection{Hamiltonian model results\label{subA}}

\begin{figure*}
\includegraphics[width=0.9\textwidth]{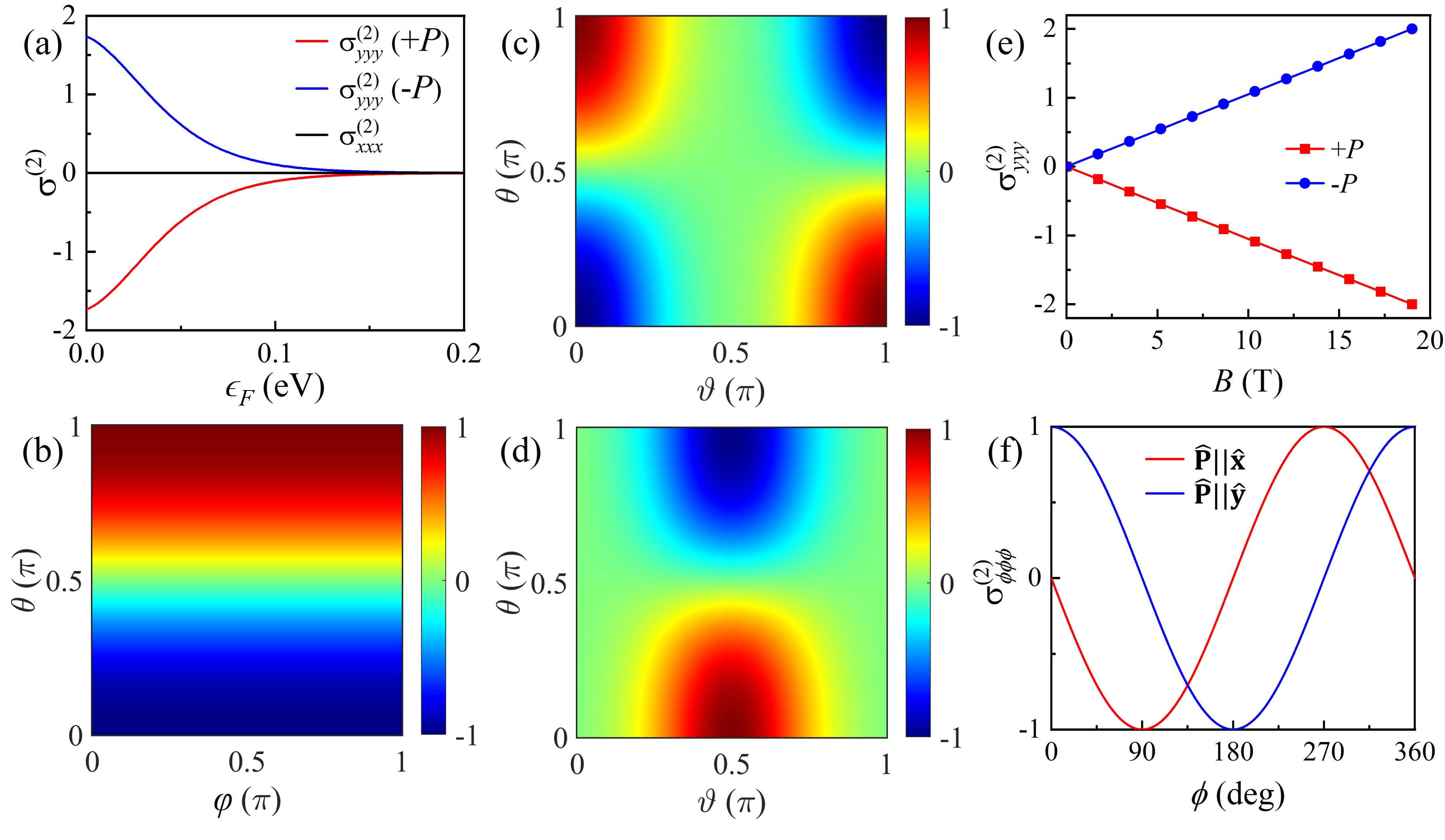}%
\caption{\label{f-2} (a) The nonlinear conductivity $\sigma^{(2)}_{xxx}$ and $\sigma^{(2)}_{yyy}$ (unit: $10^{-3}$ $e^3\tau^2|\alpha|/\hbar^3$) for $\hat{\mathbf{P}}||\hat{\mathbf{x}}$ and $B=10$ T directed along the $z$ direction as a function of the Fermi energy $\epsilon_F$. The symbols $+P$ and $-P$ denote the positive and negative polarization states, respectively. Normalized $\sigma^{(2)}_{yyy}$ at $\epsilon_F=0.1$ eV and $B=10$ T as a function of (b) ($\theta$, $\varphi$) with $\vartheta=0$ and (c) ($\theta$, $\vartheta$) with $\varphi=0$. (d) Normalized $\sigma^{(2)}_{xxx}$ at $\epsilon_F=0.1$ eV and $B=10$ T as a function of ($\theta$, $\vartheta$) with $\varphi=0$. (e) $\sigma^{(2)}_{yyy}$ at $\epsilon_F=0.1$ eV and $\hat{\mathbf{P}}||\hat{\mathbf{x}}$ as a function of $B$ directed along the $z$ direction. (f) Normalized $\sigma^{(2)}_{\phi\phi\phi}$ at $\epsilon_F=0.1$ eV and $B=10$ T directed along the $z$ direction as a function of $\phi$ for $\hat{\mathbf{P}}||\hat{\mathbf{x}}$ and $\hat{\mathbf{P}}||\hat{\mathbf{y}}$. The other parameters are assumed to be $m=0.5$ $m_0$ ($m_0$ for electron rest mass), $|\alpha|=0.5$ eV {\AA} and $T=300$ K in the Fermi distribution function.}
\end{figure*}

\begin{figure*}
\includegraphics[width=0.9\textwidth]{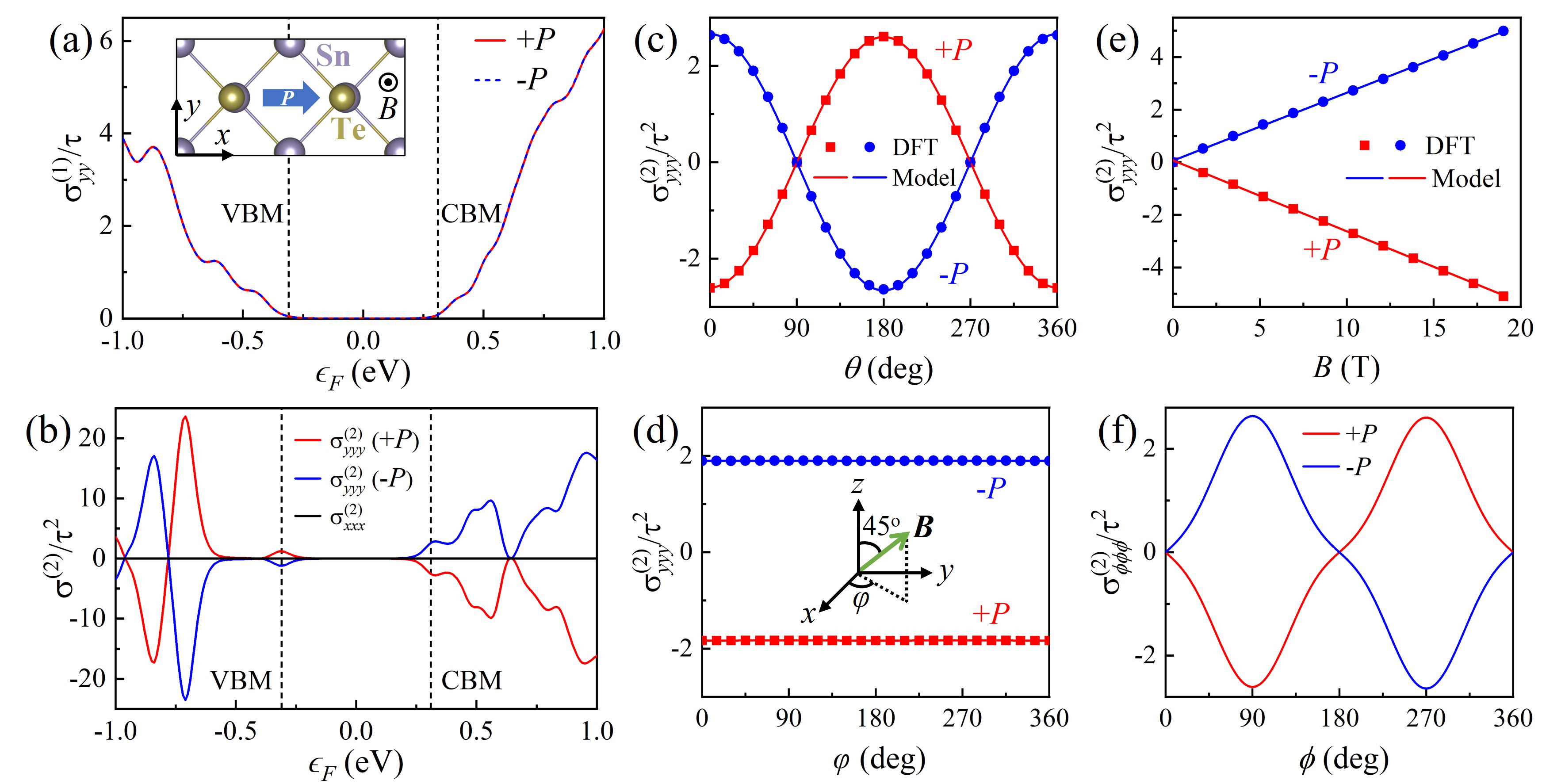}%
\caption{\label{f-3} The linear $\sigma^{(1)}_{yy}$ (unit: $10^{11}$ $\Omega^{-1}$s$^{-1}$) (a) and nonlinear conductivities $\sigma^{(2)}_{xxx}$, $\sigma^{(2)}_{yyy}$ (unit: $10^{14}$ $\Omega^{-1}$V$^{-1}$s$^{-2}$m) (b) for the SnTe monolayer at $B=10$ T directed along the $z$ direction as a function of the Fermi energy $\epsilon_F$ for positive ($+P$) and negative ($-P$) polarization states. The two vertical dashed lines denote the positions of valence band maximum (VBM) and conduction band minimum (CBM). Inset in (a) shows the atomic structure of the SnTe monolayer with polarization along the $x$ direction. $\sigma^{(2)}_{yyy}$ at $\epsilon_F=0.4$ eV and  $B=10$ T as a function of (c) $\theta$ with $\varphi=0$ and (d) $\varphi$ with $\theta=\pi/4$. (e) $\sigma^{(2)}_{yyy}$ at $\epsilon_F=0.4$ eV as a function of $B$ directed along the $z$ direction. (f) $\sigma^{(2)}_{\phi\phi\phi}$ at $\epsilon_F=0.4$ eV and $B=10$ T directed along the $z$ direction as a function of $\phi$. In (c) and (e), the symbols are DFT results while the solid lines are fits to the calculated data.}
\end{figure*}

We start from the following Hamiltonian describing the 2D ferroelectrics with SOC and Zeeman effects:
\begin{equation}\label{eq-5}
    \mathcal{H}=\frac{\hbar^2k^2}{2m}+\alpha(\hat{\mathbf{P}}\times\mathbf{k})\cdot\mathbf{\sigma}-\mu_B\mathbf{B}\cdot\mathbf{\sigma}.
\end{equation}
The first term represents the kinetic energy, where $m$ is electron effective mass (isotropic approximation), $\hbar$ the reduced Planck's constant and $\mathbf{k}=(k_x, k_y)=k(\text{cos}\phi, \text{sin}\phi)$ the wave vector given in the Cartesian and polar coordinates ($\phi$ for azimuthal angle). The second term describes the intrinsic SOC\cite{jpd113001,apl022411}, where $\alpha$ is the SOC parameter, $\hat{\mathbf{P}}=(\cos\vartheta, \sin\vartheta, 0)$ ($\vartheta$ for azimuthal angle, see Fig. \ref{f-1}) the unit vector of polarization $\mathbf{P}$ and $\mathbf{\sigma}=(\sigma_x, \sigma_y, \sigma_z)$ the vector of Pauli matrices. The third term is the Zeeman term given in terms of the Bohr magneton $\mu_B$ and the external magnetic field $\mathbf{B}$ directed along $\hat{\mathbf{B}}=(\sin\theta\cos\varphi, \sin\theta\sin\varphi, \cos\theta)$ ($\theta$ for polar angle and $\varphi$ for azimuthal angle, see Fig. \ref{f-1}) while the effective Land\'e factor is assumed as $g=2$. The eigenvalues $\epsilon_{\mathbf{k}s}$ ($s=\pm1$ for spin index) can be obtained as
\begin{equation}\label{eq-6}
  \epsilon_{\mathbf{k}s}=\frac{\hbar^2k^2}{2m}+s|\alpha(\hat{\mathbf{P}}\times\mathbf{k})-\mu_B\mathbf{B}|.
\end{equation}
For a given $\hat{\mathbf{P}}$ and $\mathbf{B}$, one first calculates $\epsilon_{\mathbf{k}s}$ from Eq. \ref{eq-6}. Then, $\sigma^{(2)}_{abc}$ can be obtained by plugging $\epsilon_{\mathbf{k}s}$ into Eq. \ref{eq-3}, where the band index $n$ is replaced by the spin index $s$. Finally, one obtains $\sigma^{(2)}_{\phi\phi\phi}$ from Eq. \ref{eq-4}.

As an illustration, we consider the polarization parallel to the $x$ axis, that is, $\hat{\mathbf{P}}||\hat{\mathbf{x}}$. Figure \ref{f-2}(a) shows the numerically calculated $\sigma^{(2)}_{xxx}$ and $\sigma^{(2)}_{yyy}$ at $B=10$ T directed along the $z$ direction as a function of the Fermi energy $\epsilon_F$ for positive ($+P$) and negative ($-P$) polarization states. Such opposite polarization states is modeled by the opposite signs of the SOC parameter $\alpha$ due to the fact that the polarity of $\alpha$ is switchable by the polarization\cite{jpd113001}. As can be seen, the magnitude of $\sigma^{(2)}_{yyy}$ decreases monotonically with increasing $\epsilon_F$ while $\sigma^{(2)}_{xxx}$ is null. This can be understood from the band dispersion Eq. \ref{eq-6}. For $\hat{\mathbf{P}}||\hat{\mathbf{x}}$ and $\hat{\mathbf{B}}||\hat{\mathbf{z}}$, we have $\epsilon_{\mathbf{k}s}=\hbar^2k^2/(2m)+s|\alpha k_y-\mu_BB|$, which suggests that $\epsilon_s(+k_x)=\epsilon_s(-k_x)$ and $\epsilon_s(+k_y)\neq\epsilon_s(-k_y)$. Thus, $\sigma^{(2)}_{yyy}$ ($\sigma^{(2)}_{xxx}$) is symmetry allowed (forbidden). In addition, the polarity of $\sigma^{(2)}_{yyy}$ is reversed upon polarization switching, indicating the feasibility of nonlinear detection of the polarization state. This is expected since the nonlinear conductivity $\sigma^{(2)}$ is a $\mathcal{P}$-odd quantity as seen from Eq. \ref{f-3}. 

Figure \ref{f-2}(b) shows the normalized $\sigma^{(2)}_{yyy}$ as a function of $\hat{\mathbf{B}}=(\sin\theta\cos\varphi, \sin\theta\sin\varphi, \cos\theta)$ for $B=10$ T and $\hat{\mathbf{P}}||\hat{\mathbf{x}}$. It is observed that $\sigma^{(2)}_{yyy}$ reveals a cosine dependent on $\theta$ and a $\varphi$ independent, that is, $\sigma^{(2)}_{yyy}\sim\cos\theta$, which is in exact accordance with the angle dependence of the second-harmonic longitudinal resistance observed in the polar metal BaMnSb$_2$\cite{prr013041}. Since $\sigma^{(2)}_{yyy}$ is caused by the band asymmetry along the $y$ direction, $\sigma^{(2)}_{yyy}\sim\cos\theta$ can be understood qualitatively from $\epsilon_{\mathbf{k}s}$. For $\hat{\mathbf{P}}||\hat{\mathbf{x}}$, we have $\epsilon_{\mathbf{k}s}=\hbar^2k^2/(2m)+s\sqrt{(\alpha k_y-\mu_BB\cos\theta)^2+(\mu_BB\sin\theta)^2}$. It is evident that the band asymmetry along the $y$ direction is dominated by the term $\alpha k_y-\mu_BB\cos\theta$. To first order in $\mu_BB$, we have $\sigma^{(2)}_{yyy}\sim\int(\partial\epsilon_{\mathbf{k}s}/\partial k_y)(\partial^2\epsilon_{\mathbf{k}s}/\partial k^2_y)dk_xdk_y\sim\int(\alpha k_y-\mu_BB\cos\theta)dk_xdk_y\sim\cos\theta$, which yields the relation $\sigma^{(2)}_{yyy}\sim\cos\theta$. Figure \ref{f-2}(c) and (d) shows the normalized $\sigma^{(2)}_{yyy}$ and $\sigma^{(2)}_{xxx}$ as a function of $\theta$ and $\vartheta$, respectively. It is seen that $\sigma^{(2)}_{yyy}\sim\cos\vartheta\cos\theta$ and $\sigma^{(2)}_{xxx}\sim\sin\vartheta\cos\theta$. Again, this is expected from the band dispersion $\epsilon_{\mathbf{k}s}=\hbar^2k^2/(2m)+s\sqrt{(\alpha k_y\cos\vartheta-\alpha k_x\sin\vartheta-\mu_BB\cos\theta)^2+(\mu_BB\sin\theta)^2}$, which indicates that the band asymmetry is dominated by the terms $\alpha k_y\cos\vartheta-\mu_BB\cos\theta$ and $-\alpha k_x\sin\vartheta-\mu_BB\cos\theta$ along $k_y$ and $k_x$ directions, respectively. Thus, to first order in $\mu_BB$, we have $\sigma^{(2)}_{yyy}\sim\int(\partial\epsilon_{\mathbf{k}s}/\partial k_y)(\partial^2\epsilon_{\mathbf{k}s}/\partial k^2_y)dk_xdk_y\sim\int\cos\vartheta(\alpha k_y\cos\vartheta-\alpha k_x\sin\vartheta-\mu_BB\cos\theta)dk_xdk_y\sim\cos\vartheta\cos\theta$ and $\sigma^{(2)}_{xxx}\sim\int(\partial\epsilon_{\mathbf{k}s}/\partial k_x)(\partial^2\epsilon_{\mathbf{k}s}/\partial k^2_x)dk_xdk_y\sim\int\sin\vartheta(\alpha k_y\cos\vartheta-\alpha k_x\sin\vartheta-\mu_BB\cos\theta)dk_xdk_y\sim\sin\vartheta\cos\theta$. We shall derive the analytical formulas for $\sigma^{(2)}$ exactly in the following.

Figure \ref{f-2}(e) shows $\sigma^{(2)}_{yyy}$ for $\hat{\mathbf{P}}||\hat{\mathbf{x}}$ as a function of $B$ directed along the $z$ direction. We can see $\sigma^{(2)}_{yyy}$ increases linearly with $B$. Figure \ref{f-2}(f) displays the normalized $\sigma^{(2)}_{\phi\phi\phi}$ at $\epsilon_F=0.1$ eV and $B=10$ T directed along the $z$ direction as a function of $\phi$. For both polarization states, $\sigma^{(2)}_{\phi\phi\phi}$ reveals the significant spatial anisotropy characterized by the maximum or zero nonlinear conductivity for certain $\phi$'s. In addition, $\sigma^{(2)}_{\phi\phi\phi}$ reveals a sine (cosine) dependent on $\phi$ for $\hat{\mathbf{P}}||\hat{\mathbf{x}}$ ($\hat{\mathbf{P}}||\hat{\mathbf{y}}$), that is, $\sigma^{(2)}_{\phi\phi\phi}\sim\sin\phi$ for $\hat{\mathbf{P}}||\hat{\mathbf{x}}$ and $\sigma^{(2)}_{\phi\phi\phi}\sim\cos\phi$ for $\hat{\mathbf{P}}||\hat{\mathbf{y}}$.

We see that $\sigma^{(2)}$ can be significantly tuned by the polarization and magnetic field. We now derive the analytical formulas for $\sigma^{(2)}$ to explain the above observed nonlinear transport phenomena. By using the Gauss's theorem, the integral in Eq. \ref{eq-3} can be transformed into the Fermi contour. In a zero-temperature limit, that is, $\partial f_n/\partial\epsilon_{\mathbf{k}}=-\delta(\epsilon_{n\mathbf{k}}-\epsilon_F)$, we may rewrite Eq. \ref{eq-3} in the form
\begin{equation}\label{eq-7}
    \sigma^{(2)}_{abc}=-\frac{e^3\tau^2}{4\pi^2\hbar^3}\sum_n\int^{2\pi}_0 [\frac{\partial\epsilon_{n\mathbf{k}}}{\partial k_a}\frac{\partial^2\epsilon_{n\mathbf{k}}}{\partial k_b\partial k_b}\frac{k}{\mid\nabla_{\mathbf{k}}\epsilon_{n\mathbf{k}}\mid}]_{\mathbf{k}=\mathbf{k}_{F}}d\phi,
\end{equation}
where $\mathbf{k}_{F}$ is the Fermi wave vector. It is noteworthy that the Zeeman energy ($\sim0.1$ meV/T) is rather small for applied magnetic fields of a few Tesla or less. It is thus legitimate to assume the weak-field or high-density regime, namely $|\alpha| k_F\gg\mu_BB$. To first order in $\mu_BB$, we obtain
\begin{equation}\label{eq-8}
\sigma^{(2)}_{xxx}\approx\sigma_0\sin\vartheta\cos\theta,
\end{equation}
and
\begin{equation}\label{eq-9}
\sigma^{(2)}_{yyy}\approx-\sigma_0\cos\vartheta\cos\theta,
\end{equation}
and
\begin{equation}\label{eq-10}
\sigma^{(2)}_{\phi\phi\phi}\approx-\sigma_0\sin(\phi-\vartheta)\cos\theta,
\end{equation}
where $\sigma_0$ is given by
\begin{equation}\label{eq-11}
\sigma_0=\frac{e^3\tau^2m\alpha^3\mu_BB}{8\sqrt{2}\pi\hbar^2\sqrt{\epsilon_F}(m\alpha^2+2\hbar^2\epsilon_F)^{\frac{3}{2}}}.
\end{equation}
Details of the derivation of Eqs. \ref{eq-8}-\ref{eq-11} are presented in the Appendix. We find that Eqs. \ref{eq-8}-\ref{eq-11} are in exact accordance with numerical results shown in Fig. \ref{f-2}.

\subsection{DFT results for the SnTe monolayer\label{subB}}

Having demonstrated the magnetic field and polarization tunable anisotropic nonlinear conductivity based on the general Hamiltonian model. We next exemplify those phenomena in the 2D ferroelectric SnTe monolayer based on the DFT calculations.

A SnTe monolayer (space group $Pmn2_1$) crystallizes in the phosphorene structure and the in-plane ferroelectric polarization is induced by the distortion between Sn and Te atoms, as shown in the inset of Fig. \ref{f-3}(a). In addition, it is a narrow gap ($\sim0.6$ eV) semiconductor and reveals high Curie temperature of $270 $ K\cite{science274} as well as sizable SOC of $\sim2.4$ eV Å\cite{apl022411}. Our calculated lattice constants of $a=4.587$, $b=4.569$ Å and the polarization of $1.34\times10^{-10}$ C/m are in good accordance with previous results\cite{apl132904,2d025026,prl076801,nc3965,npj61}. We first examine the polarization tunable conductivity. As shown in Fig. \ref{f-3}(a), the linear conductivity $\sigma^{(1)}_{yy}$ (also $\sigma^{(1)}_{xx}$, not shown) is independent of polarization direction. This is legitimate since the linear conductivity $\sigma^{(1)}$ is a $\mathcal{P}$-even quantity. On the contrary, as shown in Fig. \ref{f-3}(b), the nonlinear conductivity $\sigma^{(2)}_{yyy}$ is switchable with polarization reversal indicative of nonlinear detection of the polarization state. By using the polarization switchable $\sigma^{(2)}_{yyy}$, one could electrically control the ``forward direction" of rectifying devices. Specifically, according to Eq. \ref{f-1}, we have $|J_{+y}(+P)|<|J_{-y}(+P)|$ and $|J_{+y}(-P)|>|J_{-y}(-P)|$. In addition, $\sigma^{(2)}_{xxx}$ is null consistent with model results [see Fig. \ref{f-2}(a)] and the following symmetry arguments.

We now turn to the magnetic field tunable nonlinear conductivity. Figure \ref{f-3}(c) displays $\sigma^{(2)}_{yyy}$ at $\epsilon_F=0.4$ eV ($\sim0.1$ eV above CBM) and $B=10$ T as a function of $\theta$ with $\varphi=0$. As we can see, the DFT calculated data are nicely fitted by a cosine curve i.e. the formula Eq. \ref{eq-9}. Moreover, as shown in Fig. \ref{f-3}(d), $\sigma^{(2)}_{yyy}$ at a fixed $\theta$'s is $\varphi$ independent as consistent with model Eq. \ref{eq-9}. In Fig. \ref{f-3}(e), we plot $\sigma^{(2)}_{yyy}$ as a function of magnetic field $B$ directed along the $z$ direction. A clear linear increase with $B$ is observed as similar to model results [see Fig. \ref{f-2}(e)] and consistent with the derived analytic formula Eq. \ref{eq-11}.

Finally, the anisotropic nonlinear conductivity $\sigma^{(2)}_{\phi\phi\phi}$ can be obtained from Eq. \ref{eq-4}. We first determine the symmetry allowed nonlinear conductivity components from symmetry arguments. For the SnTe monolayer with the polarization along the $x$ direction and the applied magnetic field along the $z$ direction, the magnetic point group is identified as $m'm2'$. It contains the identity $E$, the mirror reflection about the $z=0$ plane $M_z$, the twofold rotation around the $x$ axis $C_{2x}$ followed by the time reversal $\mathcal{T}$, that is, $\mathcal{T}C_{2x}$ and the mirror reflection $M_y$ followed by $\mathcal{T}$, that is, $\mathcal{T}M_{y}$. It follows from Eq. \ref{eq-3} that the transformation rules of $\sigma^{(2)}_{abc}$ under symmetry operations are equivalent to that of $k_ak_bk_c$. Thus, the symmetry operation $\mathcal{T}C_{2x}$ enforces $\sigma^{(2)}_{xxx}=\sigma^{(2)}_{xyy}=\sigma^{(2)}_{yxy}=\sigma^{(2)}_{yyx}=0$. We substitute in Eq. \ref{eq-4} and find that
\begin{equation}\label{eq-12}
\sigma^{(2)}_{\phi\phi\phi}=\sigma^{(2)}_{yyy}\sin^3\phi+(2\sigma^{(2)}_{xxy}+\sigma^{(2)}_{yxx})\sin\phi\cos^2\phi.
\end{equation}
As shown in Fig. \ref{f-3}(f), $\sigma^{(2)}_{\phi\phi\phi}$ is significantly anisotropic and its polarity is locked to the polarization state.

In addition to the SnTe, other group-IV tellurides such as SiTe and GeTe have the same crystal structures and sizable SOCs as that of the SnTe\cite{apl132904}. Thus, the similar polarization and magnetic field control of anisotropic nonlinear transport is expected in other group-IV tellurides.

\section{Discussion and conclusions\label{sec4}}

In this work, we consider the longitudinal nonlinear response from the nonlinear Drude conductivity caused by a band asymmetry. As mentioned above, the intrinsic quantum metric also contributes to the longitudinal nonlinear response. Since they have distinct relaxation time dependencies, it is feasible to distinguish the Drude and quantum metric weights from the scaling behavior with respect to the square of the linear conductivity, as demonstrated in experiment\cite{na487}. It is enlightening to examine quantum metric-induced longitudinal nonlinear transport in 2D ferroelectrics. Second, we calculate the nonlinear Drude conductivity based on the relaxation time $\tau$ approximation and the extrinsic contributions beyond $\tau$ approximation\cite{prb035436} are neglected. Lastly, it is instructive to explore the nonlinear transport in the bulk ferroelectrics with sizable SOC and the theoretical formalism to analyze nonlinear transport phenomena can be readily generalized to bulk ferroelectrics.

In summary, using the Boltzmann transport theory, we studied the polarization and magnetic field tunable anisotropic nonlinear transport in 2D ferroelectrics with in-plane polarization based on the general model and DFT calculations. We show that the nonlinear conductivity can be significantly tuned by the polarization and an external magnetic field. In particular, we derive the analytical formulas for the nonlinear conductivity, which are in good accordance with numerical and DFT results. We further exemplify those phenomena in the 2D ferroelectric SnTe monolayer. Our results are expected to enrich the nonlinear transport physics in 2D ferroelectrics and  open avenues to design future rectifying devices.

\begin{center}
{\bf ACKNOWLEDGMENTS}
\end{center}

The authors thank Yuhao Fu and Weizhao Chen for helpful discussions. This research was supported by the National Natural Science Foundation of China (Grant No. 12274102) and the Fundamental Research Funds for the Central Universities (Grant No. FRFCU5710053421, No. HIT.OCEF.2023031). The atomic structures were produced using the VESTA software\cite{vesta}.

\begin{center}
{\bf DATA AVAILABILITY}
\end{center}

The data that support the findings of this article are not publicly available upon publication because it is not technically feasible and/or the cost of preparing, depositing, and hosting the data would be prohibitive within the terms of this research project. The data are available from the authors upon reasonable request.

\begin{widetext}
\begin{center}
{\bf APPENDIX: DERIVATION OF EQS. \ref{eq-8}-\ref{eq-11}}
\end{center}
From Eq. \ref{eq-6}, we have $\epsilon_{\mathbf{k}s}=\hbar^2k^2/(2m)+s\sqrt{\alpha^2k^2\sin^2(\phi-\vartheta)-2\alpha\mu_BBk\cos\theta\sin(\phi-\vartheta)+\mu^2_BB^2}$. To first order in $\mu_BB$, we find that
\begin{equation}\label{aq-1}
\renewcommand{\theequation}{A1}
\begin{aligned}
&\frac{\partial\epsilon_{\mathbf{k}s}}{\partial k_x}=\frac{\hbar^2k_x}{m}-s\frac{\alpha\sin\vartheta[\alpha k\sin(\phi-\vartheta)-\mu_BB\cos\theta]}{\sqrt{\alpha^2k^2\sin^2(\phi-\vartheta)-2\alpha\mu_BBk\cos\theta\sin(\phi-\vartheta)+\mu^2_BB^2}}\approx\frac{\hbar^2k_x}{m}-s|\alpha|\sin\vartheta\text{sgn}[\sin(\phi-\vartheta)],\\
&\frac{\partial\epsilon_{\mathbf{k}s}}{\partial k_y}=\frac{\hbar^2k_y}{m}+s\frac{\alpha\cos\vartheta[\alpha k\sin(\phi-\vartheta)-\mu_BB\cos\theta]}{\sqrt{\alpha^2k^2\sin^2(\phi-\vartheta)-2\alpha\mu_BBk\cos\theta\sin(\phi-\vartheta)+\mu^2_BB^2}}\approx\frac{\hbar^2k_y}{m}+s|\alpha|\cos\vartheta\text{sgn}[\sin(\phi-\vartheta)],\\
&\frac{\partial^2\epsilon_{\mathbf{k}s}}{\partial k_x^2}=\frac{\hbar^2}{m}+s\frac{\alpha^2\mu^2_BB^2\sin^2\vartheta\sin^2\theta}{[\alpha^2k^2\sin^2(\phi-\vartheta)-2\alpha\mu_BBk\cos\theta\sin(\phi-\vartheta)+\mu^2_BB^2]^{\frac{3}{2}}}\approx\frac{\hbar^2}{m},\\
&\frac{\partial^2\epsilon_{\mathbf{k}s}}{\partial k_y^2}=\frac{\hbar^2}{m}+s\frac{\alpha^2\mu^2_BB^2\cos^2\vartheta\sin^2\theta}{[\alpha^2k^2\sin^2(\phi-\vartheta)-2\alpha\mu_BBk\cos\theta\sin(\phi-\vartheta)+\mu^2_BB^2]^{\frac{3}{2}}}\approx\frac{\hbar^2}{m},\\
&\frac{\partial^2\epsilon_{\mathbf{k}s}}{\partial k_y\partial k_x}=\frac{\partial^2\epsilon_{\mathbf{k}s}}{\partial k_x\partial k_y}=-s\frac{\alpha^2\mu^2_BB^2\sin\vartheta\cos\vartheta\sin^2\theta}{[\alpha^2k^2\sin^2(\phi-\vartheta)-2\alpha\mu_BBk\cos\theta\sin(\phi-\vartheta)+\mu^2_BB^2]^{\frac{3}{2}}}\approx0.
\end{aligned}
\end{equation}
From $\epsilon_{\mathbf{k}s}$, the Fermi wave number $k_{Fs}$ can be obtained as $k_{Fs}\approx-sk_\phi+\sqrt{k_\phi^2+k_0^2+s\frac{2\mu_BB\cos\theta k_\phi}{\alpha\sin(\phi-\vartheta)}}$, where $k_0=\sqrt{2m\epsilon_F/\hbar^2}$ and $k_\phi=m|\alpha\sin(\phi-\vartheta)|/\hbar^2$. To first order in $\mu_BB$, the absolute value of $\nabla_{\mathbf{k}}\epsilon_{\mathbf{k}s}$ takes the form
\begin{equation}\label{aq-2}
\renewcommand{\theequation}{A2}
\begin{aligned}
&|\nabla_{\mathbf{k}}\epsilon_{\mathbf{k}s}|=|\frac{\hbar^2\mathbf{k}}{m}+s\frac{\alpha[\alpha k\sin(\phi-\vartheta)-\mu_BB\cos\theta](\cos\vartheta\hat{\mathbf{y}}-\sin\vartheta\hat{\mathbf{x}})}{\sqrt{\alpha^2k^2\sin^2(\phi-\vartheta)-2\alpha\mu_BBk\cos\theta\sin(\phi-\vartheta)+\mu^2_BB^2}}|\\
&=\sqrt{\frac{\hbar^4k^2}{m^2}+s\frac{2\hbar^2\alpha k[\alpha k\sin(\phi-\vartheta)-\mu_BB\cos\theta]\sin(\phi-\vartheta)}{m\sqrt{\alpha^2k^2\sin^2(\phi-\vartheta)-2\alpha\mu_BBk\cos\theta\sin(\phi-\vartheta)+\mu^2_BB^2}}+\frac{\alpha^2 [\alpha k\sin(\phi-\vartheta)-\mu_BB\cos\theta]^2}{\alpha^2k^2\sin^2(\phi-\vartheta)-2\alpha\mu_BBk\cos\theta\sin(\phi-\vartheta)+\mu^2_BB^2}}\\
&\approx\sqrt{\frac{\hbar^4k^2}{m^2}+s\frac{2\hbar^2k|\alpha\sin(\phi-\vartheta)|}{m}+\alpha^2}.
\end{aligned}
\end{equation}
At the Fermi energy, $\mathbf{k}=\mathbf{k}_{Fs}$, we obtain
\begin{equation}\label{aq-3}
\renewcommand{\theequation}{A3}
\begin{aligned}
\frac{1}{|\nabla_{\mathbf{k}_{Fs}}\epsilon_{\mathbf{k}_{Fs}}|}&=\{\frac{\hbar^4k^2_{Fs}}{m^2}+s\frac{2\hbar^2k_{Fs}|\alpha\sin(\phi-\vartheta)|}{m}+\alpha^2\}^{-\frac{1}{2}}\\
&=\{\frac{\hbar^4k^2_0}{m^2}+s\frac{2\hbar^2\mu_BB\cos\theta}{m}\text{sgn}[\alpha\sin(\phi-\vartheta)]+\alpha^2\}^{-\frac{1}{2}}\\
&\approx\frac{1}{C}-s\frac{\hbar^2\mu_BB\cos\theta}{mC^3}\text{sgn}[\alpha\sin(\phi-\vartheta)],
\end{aligned}
\end{equation}
where $C$ is defined as $C\equiv\sqrt{\alpha^2+2\hbar^2\epsilon_F/m}$. Substitution of Eqs. (A1) and (A3) in Eq. \ref{eq-7} and to first order in $\mu_BB$ yields
\begin{equation}\label{aq-4}
\renewcommand{\theequation}{A4}
\begin{aligned}
\sigma^{(2)}_{xxx}&=-\frac{e^3\tau^2}{4\pi^2\hbar^3}\sum_s\int^{2\pi}_0\frac{\hbar^2}{m}\{\frac{\hbar^2k_{Fs}\cos\phi}{m}-s|\alpha|\sin\vartheta\text{sgn}[\sin(\phi-\vartheta)]\}\{\frac{k_{Fs}}{C}-s\frac{\hbar^2\mu_BB\cos\theta}{C^3m}\text{sgn}[\alpha\sin(\phi-\vartheta)]k_{Fs}\}d\phi\\
&=-\frac{e^3\tau^2}{4\pi^2\hbar^3}\sum_s\int^{2\pi}_0\{\frac{\hbar^4\cos\phi k^2_{Fs}}{Cm^2}-s\frac{\hbar^6\mu_BB\cos\theta\cos\phi k^2_{Fs}}{C^3m^3}\text{sgn}[\alpha\sin(\phi-\vartheta)]\\
&-s\frac{\hbar^2\sin\vartheta|\alpha|k_{Fs}}{Cm}\text{sgn}[\sin(\phi-\vartheta)]+\frac{\hbar^4\sin\vartheta\alpha\mu_BB\cos\theta k_{Fs}}{C^3m^2}\}d\phi.\\
\end{aligned}
\end{equation}
Up to linear in $\mu_BB$, the integrals in Eq. (A4) can be calculated respectively as
\begin{equation}\label{aq-5}
\renewcommand{\theequation}{A5}
\begin{aligned}
&\sum_s\int^{2\pi}_0\frac{\hbar^4\cos\phi k^2_{Fs}}{Cm^2}d\phi\approx\sum_s\int^{2\pi}_0[\frac{2\hbar^4k^2_\phi\cos\phi}{Cm^2}+\frac{\hbar^4k^2_0\cos\phi}{Cm^2}-\frac{2\hbar^4\mu_BB\cos\theta k^2_\phi\cos\phi}{Cm^2\alpha k_0\sin(\phi-\vartheta)}\\
&+s\frac{2\hbar^4\mu_BB\cos\theta k_\phi\cos\phi}{Cm^2\alpha\sin(\phi-\vartheta)}-s\frac{2\hbar^4k_0k_\phi\cos\phi}{Cm^2}-s\frac{\hbar^4k^3_\phi\cos\phi}{Cm^2k_0}]d\phi\\
&=\sum_s\int^{2\pi}_0[\frac{2\alpha^2}{C}\sin^2(\phi-\vartheta)\cos\phi-\frac{2\alpha\mu_BB\cos\theta}{C k_0}\sin(\phi-\vartheta)\cos\phi]d\phi\\
&=\frac{4\pi\alpha\mu_BB\cos\theta\sin\vartheta}{Ck_0},
\end{aligned}
\end{equation}
and
\begin{equation}\label{aq-6}
\renewcommand{\theequation}{A6}
\begin{aligned}
&\sum_s\int^{2\pi}_0-s\frac{\hbar^6\mu_BB\cos\theta\cos\phi k^2_{Fs}}{C^3m^3}\text{sgn}[\alpha\sin(\phi-\vartheta)]d\phi\\
&\approx\sum_s\int^{2\pi}_0\frac{\hbar^6\mu_BB\cos\theta}{C^3m^3}(-2sk^2_{\phi}-sk^2_0+2k_0k_{\phi}+\frac{k^3_{\phi}}{k_0})\cos\phi\text{sgn}[\alpha\sin(\phi-\vartheta)]d\phi\\
&=\sum_s\int^{2\pi}_0\frac{\mu_BB\cos\theta}{C^3}[-\frac{2\hbar^4\alpha k_0\sin\vartheta\cos^2\phi}{m^2}+\frac{\alpha^3\sin^3(\phi-\vartheta)\cos\phi}{k_0}]d\phi\\
&=-\frac{4\pi\hbar^4\alpha k_0\mu_BB\cos\theta\sin\vartheta}{C^3m^2}-\frac{3\pi\alpha^3\mu_BB\cos\theta\sin\vartheta}{2C^3k_0},
\end{aligned}
\end{equation}
and
\begin{equation}\label{aq-7}
\renewcommand{\theequation}{A7}
\begin{aligned}
&\sum_s\int^{2\pi}_0-s\frac{\hbar^2\sin\vartheta|\alpha|k_{Fs}}{Cm}\text{sgn}[\sin(\phi-\vartheta)]d\phi\\
&\approx\sum_s\int^{2\pi}_0\{\frac{\hbar^2\sin\vartheta|\alpha|k_\phi}{Cm}\text{sgn}[\sin(\phi-\vartheta)]-s\frac{\hbar^2\sin\vartheta|\alpha|k_0}{Cm}\text{sgn}[\sin(\phi-\vartheta)]d\phi\\
&-s\frac{\hbar^2\sin\vartheta|\alpha|k^2_\phi}{2Cmk_0}\text{sgn}[\sin(\phi-\vartheta)]-\frac{\hbar^2\sin\vartheta|\alpha|\mu_BB\cos\theta k_\phi}{Cm\alpha k_0|\sin(\phi-\vartheta)|}\}d\phi\\
&=\sum_s\int^{2\pi}_0-\frac{\hbar^2\sin\vartheta|\alpha|\mu_BB\cos\theta k_\phi}{Cm\alpha k_0|\sin(\phi-\vartheta)|}d\phi=-\frac{4\pi\alpha\mu_BB\cos\theta\sin\vartheta}{Ck_0},
\end{aligned}
\end{equation}
and
\begin{equation}\label{aq-8}
\renewcommand{\theequation}{A8}
\begin{aligned}
&\sum_s\int^{2\pi}_0\frac{\hbar^4\sin\vartheta\alpha\mu_BB\cos\theta k_{Fs}}{C^3m^2}d\phi\\
&\approx\sum_s\int^{2\pi}_0[-s\frac{\hbar^4\sin\vartheta\alpha\mu_BB\cos\theta k_{\phi}}{C^3m^2}+\frac{\hbar^4\sin\vartheta\alpha k_0\mu_BB\cos\theta}{C^3m^2}+\frac{\hbar^4\sin\vartheta\alpha\mu_BB\cos\theta k^2_{\phi}}{2C^3m^2k_0}]d\phi\\
&=\sum_s\int^{2\pi}_0[\frac{\hbar^4\sin\vartheta\alpha k_0\mu_BB\cos\theta}{C^3m^2}+\frac{\alpha^3\sin\vartheta\mu_BB\cos\theta}{2C^3k_0}\sin^2(\phi-\vartheta)]d\phi\\
&=\frac{4\pi\hbar^4\alpha k_0\mu_BB\cos\theta\sin\vartheta}{C^3m^2}+\frac{\pi\alpha^3\mu_BB\cos\theta\sin\vartheta}{C^3k_0}.
\end{aligned}
\end{equation}
Substitution of Eqs. (A5)-(A8) in Eq. (A4) yields
\begin{equation}\label{aq-9}
\renewcommand{\theequation}{A9}
\begin{aligned}
&\sigma^{(2)}_{xxx}\approx\frac{e^3\tau^2\alpha^3\mu_BB\cos\theta\sin\vartheta}{8\pi\hbar^3C^3k_0}=\sigma_0\sin\vartheta\cos\theta,\\
&\sigma_0\equiv\frac{e^3\tau^2\alpha^3\mu_BB}{8\pi\hbar^3C^3k_0}=\frac{e^3\tau^2m\alpha^3\mu_BB}{8\sqrt{2}\pi\hbar^2\sqrt{\epsilon_F}(m\alpha^2+2\hbar^2\epsilon_F)^{\frac{3}{2}}}.
\end{aligned}
\end{equation}
For $\sigma^{(2)}_{yyy}$, we have
\begin{equation}\label{aq-10}
\renewcommand{\theequation}{A10}
\begin{aligned}
\sigma^{(2)}_{yyy}&=-\frac{e^3\tau^2}{4\pi^2\hbar^3}\sum_s\int^{2\pi}_0\frac{\hbar^2}{m}\{\frac{\hbar^2k_{Fs}\sin\phi}{m}+s|\alpha|\cos\vartheta\text{sgn}[\sin(\phi-\vartheta)]\}\{\frac{k_{Fs}}{C}-s\frac{\hbar^2\mu_BB\cos\theta}{C^3m}\text{sgn}[\alpha\sin(\phi-\vartheta)]k_{Fs}\}d\phi\\
&=-\frac{e^3\tau^2}{4\pi^2\hbar^3}\sum_s\int^{2\pi}_0\{\frac{\hbar^4\sin\phi k^2_{Fs}}{Cm^2}-s\frac{\hbar^6\mu_BB\cos\theta\sin\phi k^2_{Fs}}{C^3m^3}\text{sgn}[\alpha\sin(\phi-\vartheta)]\\
&+s\frac{\hbar^2\cos\vartheta|\alpha|k_{Fs}}{Cm}\text{sgn}[\sin(\phi-\vartheta)]-\frac{\hbar^4\cos\vartheta\alpha\mu_BB\cos\theta k_{Fs}}{C^3m^2}\}d\phi.\\
\end{aligned}
\end{equation}
Up to linear in $\mu_BB$, the integrals in Eq. (A10) can be calculated respectively as
\begin{equation}\label{aq-11}
\renewcommand{\theequation}{A11}
\begin{aligned}
&\sum_s\int^{2\pi}_0\frac{\hbar^4\sin\phi k^2_{Fs}}{Cm^2}d\phi\approx\sum_s\int^{2\pi}_0[\frac{2\hbar^4k^2_\phi\sin\phi}{Cm^2}+\frac{\hbar^4k^2_0\sin\phi}{Cm^2}-\frac{2\hbar^4\mu_BB\cos\theta k^2_\phi\sin\phi}{Cm^2\alpha k_0\sin(\phi-\vartheta)}\\
&+s\frac{2\hbar^4\mu_BB\cos\theta k_\phi\sin\phi}{Cm^2\alpha\sin(\phi-\vartheta)}-s\frac{2\hbar^4k_0k_\phi\sin\phi}{Cm^2}-s\frac{\hbar^4k^3_\phi\sin\phi}{Cm^2k_0}]d\phi\\
&=\sum_s\int^{2\pi}_0[\frac{2\alpha^2}{C}\sin^2(\phi-\vartheta)\sin\phi-\frac{2\alpha\mu_BB\cos\theta}{C k_0}\sin(\phi-\vartheta)\sin\phi]d\phi\\
&=-\frac{4\pi\alpha\mu_BB\cos\theta\cos\vartheta}{Ck_0},
\end{aligned}
\end{equation}
and
\begin{equation}\label{aq-12}
\renewcommand{\theequation}{A12}
\begin{aligned}
&\sum_s\int^{2\pi}_0-s\frac{\hbar^6\mu_BB\cos\theta\sin\phi k^2_{Fs}}{C^3m^3}\text{sgn}[\alpha\sin(\phi-\vartheta)]d\phi\\
&\approx\sum_s\int^{2\pi}_0\frac{\hbar^6\mu_BB\cos\theta}{C^3m^3}(-2sk^2_{\phi}-sk^2_0+2k_0k_{\phi}+\frac{k^3_{\phi}}{k_0})\sin\phi\text{sgn}[\alpha\sin(\phi-\vartheta)]d\phi\\
&=\sum_s\int^{2\pi}_0\frac{\mu_BB\cos\theta}{C^3}[\frac{2\hbar^4\alpha k_0\cos\vartheta\sin^2\phi}{m^2}+\frac{\alpha^3\sin^3(\phi-\vartheta)\sin\phi}{k_0}]d\phi\\
&=\frac{4\pi\hbar^4\alpha k_0\mu_BB\cos\theta\cos\vartheta}{C^3m^2}+\frac{3\pi\alpha^3\mu_BB\cos\theta\cos\vartheta}{2C^3k_0},
\end{aligned}
\end{equation}
and
\begin{equation}\label{aq-13}
\renewcommand{\theequation}{A13}
\begin{aligned}
\sum_s\int^{2\pi}_0s\frac{\hbar^2\cos\vartheta|\alpha|k_{Fs}}{Cm}\text{sgn}[\sin(\phi-\vartheta)]d\phi\approx\frac{4\pi\alpha\mu_BB\cos\theta\cos\vartheta}{Ck_0},
\end{aligned}
\end{equation}
and
\begin{equation}\label{aq-14}
\renewcommand{\theequation}{A14}
\sum_s\int^{2\pi}_0-\frac{\hbar^4\cos\vartheta\alpha\mu_BB\cos\theta k_{Fs}}{C^3m^2}d\phi\approx-\frac{4\pi\hbar^4\alpha k_0\mu_BB\cos\theta\cos\vartheta}{C^3m^2}-\frac{\pi\alpha^3\mu_BB\cos\theta\cos\vartheta}{C^3k_0},
\end{equation}
Substitution of Eqs. (A11)-(A14) in Eq. (A10) yields
\begin{equation}\label{aq-15}
\renewcommand{\theequation}{A15}
\sigma^{(2)}_{yyy}\approx-\frac{e^3\tau^2\alpha^3\mu_BB\cos\theta\cos\vartheta}{8\pi\hbar^3C^3k_0}=-\sigma_0\cos\vartheta\cos\theta.
\end{equation}
From Eqs. \ref{eq-7} and (A1), the other components are $\sigma^{(2)}_{xyy}=\sigma^{(2)}_{xxx}$, $\sigma^{(2)}_{yxx}=\sigma^{(2)}_{yyy}$, $\sigma^{(2)}_{xxy}=\sigma^{(2)}_{xyx}=\sigma^{(2)}_{yxy}=\sigma^{(2)}_{yyx}\approx0$. We substitute in Eq. \ref{eq-4} and find that
\begin{equation}\label{aq-16}
\renewcommand{\theequation}{A16}
\begin{aligned}
\sigma^{(2)}_{\phi\phi\phi}&=\sigma^{(2)}_{xxx}\cos^3\phi+\sigma^{(2)}_{yyy}\sin^3\phi+(\sigma^{(2)}_{xyy}+2\sigma^{(2)}_{yxy})\sin^2\phi\cos\phi+(\sigma^{(2)}_{yxx}+2\sigma^{(2)}_{xxy})\sin\phi\cos^2\phi.\\
&=\sigma^{(2)}_{xxx}\cos^3\phi+\sigma^{(2)}_{yyy}\sin^3\phi+\sigma^{(2)}_{xxx}\sin^2\phi\cos\phi+\sigma^{(2)}_{yyy}\sin\phi\cos^2\phi\\
&=\sigma^{(2)}_{xxx}\cos\phi+\sigma^{(2)}_{yyy}\sin\phi\\
&=-\sigma_0\sin(\phi-\vartheta)\cos\theta.
\end{aligned}
\end{equation}

\end{widetext}

\end{document}